%
%

\input phyzzx

\pubnum{93-33}
\date{}
\titlepage
\title{\fourteenbf
$W_\infty$ coherent states and path-integral derivation of
bosonization of non-relativistic fermions in one dimension}
\author{\rm Avinash Dhar, Gautam Mandal and Spenta R. Wadia}
\address{\rm Tata Institute of Fundamental Research \break Homi Bhabha Road,
Bombay 400 005, India}
\abstract{
We complete the proof of bosonization of noninteracting nonrelativistic
fermions in one space dimension by deriving the bosonized action using
$W_\infty$ coherent  states in the fermion path-integral. This action was
earlier derived by us using the method of coadjoint orbits.  We also
discuss the classical limit of the bosonized theory and indicate the
precise nature of the truncation of the full theory that leads to the
collective field theory.}

\endpage

\REF\BLOCH{F. Bloch, ZS. Phys. 81 (1933) 363, Helv. Phys. Acta 7 (1934)
385.}

\REF\BOHM{See D. Bohm and D. Pines, Phys. Rev. 92 (1953) 609, and
references therein.}

\REF\TOMONAGA{S. Tomonaga, Prog. Theor. Phys. 5 (1950)  544.}

\REF\LUTTINGER{J.M. Luttinger, J. Math. Phys. 4 (1963) 1154.}

\REF\LIEBMATTIS{D. Mattis and E. Lieb, J. Math. Phys. 6 (1965) 375.}

\REF\LUTHER{A. Luther and I. Peschel, Phys. Rev. B9 (1974) 2911.}

\REF\HALDANE{F.D.M. Haldane, J. Phys. C 14 (1981) 2585.}

\REF\MANDELSTAM{S. Mandelstam, Phys. Rev. D11 (1985) 3026.}

\REF\SAKITA {A. Jevicki and B. Sakita, Nucl. Phys. B165 (1980) 511.}

\REF\DMWCLASSICAL{A. Dhar, G. Mandal and S.R. Wadia, Int. J. Mod. Phys. A8
(1993) 325.}

\REF\DMW {A. Dhar, G. Mandal and S.R. Wadia, Mod. Phys. Lett. A7 (1992) 3129.}

\REF\DASJEVICKI{S.R. Das and A. Jevicki, Mod. Phys. Lett. A5 (1990) 1639.}

\REF\BREZ {E. Brezin, C. Itzykson, G. Parisi and J.B. Zuber, Comm. Math.
Phys. 59 (1978) 35; E. Brezin, V.A. Kazakov and Al.B. Zamolochikov, Nucl.
Phys. B338 (1990) 673; D. Gross and N. Miljkovich, Nucl. Phys. B238 (1990)
217; P. Ginsparg and J. Zinn-Justin, Phys. Lett. B240 (1990) 333; G.,
Parisi, Europhys. Lett. 11 (1990) 595.}

\REF\SENG {A.M. Sengupta and S.R. Wadia, Int. J. Mod. Phys. A6 (1991)
1961; G. Mandal, A.M. Sengupta and S.R. Wadia, Mod. Phys. Lett. A6 (1991)
1465.}

\REF\GROS {D.J. Gross and I. Klebanov, Nucl. Phys. B352 (1990) 671.}

\REF\MOOR {G. Moore, Nucl. Phys. B368 (1992) 557.}

\REF\DAS {S.R. Das, A. Dhar, G. Mandal and S.R. Wadia, Mod. Phys. Letts.
A7 (1992) 71.}

\REF\DASTWO {S.R. Das, A. Dhar, G. Mandal and S.R. Wadia, Int. J. Mod.
Phys. A7 (1992) 5165.}

\REF\PERE {A. Perelomov, Generalized Coherent States and Their
Applications, Springer-Verlag.}

\REF\POL{J. Polchinski, Nucl. Phys. B346 (1990) 253.}

\REF\DMWTIME {A. Dhar, G. Mandal and S.R. Wadia,
Int. J. Mod. Phys. A8 (1993) 3811.}

\noindent 1.{\bf Introduction and Summary}

Bosonization of non-relativistic fermions is an important problem with a
long history. It was observed by Bloch [\BLOCH] many years ago
in his famous
calculation of the stopping power of charged particles that the low energy
excitations of a fermi gas can be described more suitably
(within certain approximations) in terms of density fluctuations
of the fermi gas (``sound waves'') rather than in terms of individual excited
particles and holes.  He also observed that if the sound waves are
quantized the quanta obey bose statistics under these approximations. In
these treatments the fermions were considered to be basically free. Bohm
{\it et al} [\BOHM]
considered the effect of Coulomb interactions between the fermions and
found a new kind of collective oscillation (``plasma oscillation'')  which
had a characteristic frequency independent of the wave-number for low
wave-numbers. The corresponding quanta (plasmons) again were found to be
bosonic under some approximations.  Tomonaga wrote a comprehensive article
[\TOMONAGA]
in which he showed the formal equivalence of the low energy sector of a
system of free non-relativistic fermions with that of a  free
relativistic boson in
the case of one dimension under a set of well-defined approximations. The
approximations basically consisted of (1) considering only those states of
the fermi theory which are built from holes or excited particles
(either left-moving or right-moving) which have  wave-numbers
between $3k_F/4$ and $5k_F/4$, where $k_F$ is the magnitude of the
wave-number for the fermi surface, and (2) ignoring sound quanta which
have  wave-numbers greater than $k_F/2$.
The sound quanta under these
approximations were the same as quantized density waves in the fermi gas.
Tomonaga was also able to incorporate the effect of interactions under
this scheme, exploiting the beautiful observation that interactions
between fermions, which typically require four-fermion terms in the
hamiltonian, involve quadratic (density-density) terms in the bose
hamiltonian, thus keeping the theory linear. The effect of interactions
was thus incorporated in the Bose language in terms of  a simple
rediagonalization of the Bose field (a Bogoliubov transformation); the
plasma oscillations for instance are trivially deduced in this way.

The question of a complete bosonization  has subsequently been addressed
by many authors. Except in the case of the Luttinger model
[\LUTTINGER]
where the
fermion has a linear dispersion relation, a complete bosonization of
non-relativistic fermions has always been elusive, even in one dimension.
The bosonization of relativistic fermions is similar in spirit to
Tomonaga's approximate bosonization because the dispersion relation in
the low energy band around the fermi surface is approximately linear.
Works on bosonization of relativistic fermions include Lieb and Mattis
[\LIEBMATTIS],
Luther and Peschel [\LUTHER], Haldane [\HALDANE]
and, from the field theory point of view,
Mandelstam [\MANDELSTAM].
A different approach to approximate bosonization
of non-relativistic fermions was taken by Jevicki and Sakita
[\SAKITA]
who exploited
the equivalence of the fermion problem (in one dimension) to matrix
models and used the method of collective variables.

A clue to the full solution of the bosonization problem can be obtained
by looking at the semi-classical picture of a fermi gas, which describes
the states of the fermi theory in terms of a fermi fluid of various shapes
(with the same area as the ground-state configuration of the fluid, if we
insist on fermion number conservation). This fermi fluid exists in the
two-dimensional phase space of the single fermion. We see therefore that
in this semi-classical approximation changes of the state of the fermi theory
correspond to area-preserving shape changes of the fermi fluid. This is
similar in spirit to Bloch's observation mentioned in the last paragraph
except that we are now talking about fluctuations of the phase space
density rather than that of the ordinary density. This classical picture
has been elaborated in [\DMWCLASSICAL].
In [\DMW]  we extended this
bosonization in terms of the phase space density to the quantum theory. In
this paper we present a first principles proof of the bosonization starting
from the fermion path integral using the techniques of coherent states.
We will also present a brief discussion of the precise nature of the
truncation of the bosonized theory  that leads to collective field theory
[\SAKITA, \DASJEVICKI].

\noindent 2. {\bf Derivation of the Path-Integral using Coherent States}

In the following we will consider the specific example of the fermion field
theory which emerges in the double scaling limit of the $c=1$ matrix model.
The discussion is however easily generalized to other one-dimensional
fermi systems.

It is well-known [\BREZ-\MOOR] that the $c=1$ matrix model is described by
the field theory of noninteracting nonrelativistic fermions in one space
dimension, defined by the action
$$
S = \int^{+\infty}_{-\infty} dt \int^{+\infty}_{-\infty} dx ~\psi^+ (x,t)
(i\partial_t - h_x) \psi(x,t)
\eqn\one
$$
where the single-particle hamiltonian $h$ is given by
$$
\eqalign{&
h_x = {1\over2} \left(-\partial^2_x + V(x)\right) \cr &
V(x) = -x^2 + {g_3 \over \sqrt{N}} x^3 + \cdots \cr &
N = \int^{+\infty}_{-\infty} dx ~\psi^+ (x,t) \psi(x,t)}
\eqn\two
$$
In the above we have chosen the zeros of energy and $x$-axis such that the
(quadratic) maximum of the potential occurs at $x=0$ and that $V_{max} =
V(0) = 0$.  The continuum (double scaling) limit is obtained by letting $N
\rightarrow \infty$ and the bare fermi energy $\epsilon_F \rightarrow 0$
while keeping the renormalized fermi energy (measured from the top of the
potential) $\mu \sim N\epsilon_F$ fixed.  The string coupling $g_{str}$ is
then given by $g_{str} \sim {1 \over |\mu|}$.  ($\mu$ is negative in our
conventions.)

In a previous work [\DMW] we presented a bosonization of \one\ using the
method of coadjoint orbits of $W_\infty$.  A heuristic derivation of this
action, starting from the fermion path integral, was also discussed
previously in [\DAS].  Here we will complete the proof of
bosonization of \one\ by deriving the boson action of ref. [\DMW] using
the method of coherent states of $W_\infty$ in the fermion path-integral.

The use of $W_\infty$ coherent states in the fermion path-integral is made
possible by the observation [\DAS] that the bosonized problem is analogous
to that of a spin in a magnetic field.  Let us recall this analogy here.
The most general (elementary) boson operator is the fermion bilocal
$$
\phi (x,y,t) \equiv \psi^+ (x,t) \psi(y,t)
\eqn\three
$$
By virtue of the fermion anticommutation relation, $\phi$ satisfies the
closed operator algebra
$$
\eqalign{
[\phi(x,y,t),\phi(x',y',t')] = \phi(x,&y',t) \delta(x'-y) \cr & -
\phi(x',y,t) \delta(x-y')}
\eqn\four
$$
Also, using the fermion equation of motion one can derive an equation of
motion for $\phi$.  Introducing the compact ``matrix'' notation,
$$
\langle x|\Phi(t)|y\rangle \equiv \phi(x,y,t),
\eqn\five
$$
this equation of motion can be written as
$$
i\partial_t\Phi + [h,\Phi] = 0,
\eqn\six
$$
where the matrix elements of $h$ are given by $\langle x|h|y\rangle = h_x
\delta(x-y)$.  Equations \four\ and \six\ describe a $W_\infty$ `spin'
system, with $h$ acting like an external magnetic field.

The $W_\infty$ algebra [\DASTWO,\DMW] \four\ can be written in a more
familiar form in terms of the new operator
$$
W(\alpha,\beta,t) \equiv \int dx~e^{i\alpha x} \phi(x+\beta/2,x-\beta/2,t)
\eqn\seven
$$
which satisfies the algebra
$$
[W(\alpha,\beta,t),W(\alpha',\beta',t)] = 2\pi i \sin
{1\over2}(\alpha\beta' - \alpha'\beta) W(\alpha+\alpha',\beta+\beta',t)
\eqn\eight
$$
In the single-particle Hilbert space the $W_\infty$ algebra is generated
by all differential operators in one-dimension, i.e. by operator of the
type $\hat x^n \hat p^m$, where $[\hat x,\hat p] = i$.  A convenient basis
is given by
$$
g(\alpha,\beta) = e^{i(\alpha \hat x-\beta\hat p)}
\eqn\nine
$$
which satisfies the algebra
$$
[g(\alpha,\beta),g(\alpha',\beta')] = 2\pi i \sin {1\over2} (\alpha\beta'
- \alpha'\beta) g(\alpha+\alpha',\beta+\beta')
\eqn\ten
$$
The $W(\alpha,\beta,t)$ realize this algebra in the fermion fock space.

The vacuum of the fermion theory is easily constructed by filling the
fermi sea to a certain fermi level, which is determined by the number of
fermions.  Let us denote this vacuum state by $|F_0\rangle$.  Coherent
states are constructed by the action of the $W_\infty$ group elements on
$|F_0\rangle$ [\PERE]
$$
|F_\theta\rangle = u(\theta) |F_0\rangle, ~~~ u \in {\cal G} W_\infty
\eqn\eleven
$$
${\cal G}W_\infty$ is the Lie group corresponding to the $W_\infty$
algebra.  In terms of the generators $W(\alpha,\beta)$, $u(\theta)$ can be
parametrized as
$$
u(\theta) = \exp[i \int d\alpha ~d\beta ~W(\alpha,\beta) \theta(\alpha,\beta)]
\eqn\twelve
$$
In general, for certain functions $\theta(\alpha,\beta)$, $u(\theta)$
would leave $|F_0\rangle$ invariant.  This subset of $u(\theta)$'s clearly
forms a subgroup $H$ of ${\cal G}W_\infty$.  So, the distinct coherent
states $|F_0\rangle$ in \eleven\ are given by the elements of the coset
${\cal G}W_\infty/H$.  This coset depends on the filling of the fermi sea.
(To illustrate this point, consider the simpler case of a finite level
system instead of $W_\infty$ (e.g. one may consider $u(N)$).   Then, it is
clear for example that in the extreme case of filling of all levels, the
coset consists of a single point, the fermi vacuum.  For partial filling
there is clearly a nontrivial coset and it may be verified by the reader
in simple cases that the coset depends on the filling.)

Having specified the ${\cal G}W_\infty$ coherent states defined on the
fermi vacuum $|F_0\rangle$, let us explain their significance in
bosonizing the fermion path integral.  Firstly, let us note that we are
interested in evaluating correlation functions involving only the bilocal
boson operator $\phi(x,y,t)$ or some (fourier) transform of it.  Because
of this it is sufficient to consider intermediate states in the
path-integral from the linear span of $\{\displaystyle \prod_i
\phi(x_i,y_i) |F_0\rangle\} \equiv {\cal F}$.  These states form a
complete set and give a resolution of the identity.  On the other hand, we
may consider the linear span of the set of coherent states
$\{|F_\theta\rangle\} \equiv {\cal E}$.
Clearly, any element in the linear span
of ${\cal F}$ is in the linear span of
${\cal E}$ and vice versa.  Hence, we may
consider a resolution of the identity in terms of the coherent states,
even though they form an overcomplete set,
$$
\int d\mu (\theta) |F_\theta\rangle \langle F_\theta| = 1
\eqn\thirteen
$$

The derivation of the path-integral now rests on the evaluation of the
short time kernal
$$
K_{t+\epsilon,\epsilon} = \langle F_{\theta(t+\epsilon)} |e^{i\epsilon H}
|F_{\theta(t)}\rangle
\eqn\fourteen
$$
where $H = \int dx ~\psi^+ h_x \psi$ is the hamiltonian of the fermion
field theory and is an element of the $W_\infty$ algebra.  One may
equivalently write it in terms of the bilocal operator $\Phi$ as $H = tr(h
\Phi)$, where we have introduced the notation $tr(AB) = \int dx~dy \langle
x|A|y\rangle \langle y|B|x\rangle$.  Expanding \fourteen\ in $\epsilon$, we get
$$
K_{t+\epsilon,\epsilon} = \langle
F_{\theta(t+\epsilon)}|F_{\theta(t)}\rangle + i\epsilon \langle
F_{\theta(t)}|H|F_{\theta(t)}\rangle + O(\epsilon^2)
\eqn\fifteen
$$

To evaluate the first term in \fifteen\ we use \eleven\ and expand in
$\epsilon$.  We get
$$
\langle F_{\theta(t+\epsilon)}|F_{\theta(t)}\rangle = 1 + i\epsilon
\langle F_0|u^+(\theta(t)) i\partial_t u(\theta(t))|F_0\rangle +
O(\epsilon^2)
\eqn\sixteen
$$
The operator inside the expectation value in \sixteen\
is an element of the
$W_\infty$ algebra and so it can be expanded in the basis provided by
$W(\alpha,\beta)$,
$$
u^+(\theta(t)) i\partial_t u(\theta(t)) = \int d\alpha~d\beta~
C_{\alpha\beta} (\theta(t),\partial_t\theta(t)) W(\alpha,\beta)
\eqn\seventeen
$$
Let us now define the single-particle analogue of \twelve ,
$$
g(\theta) = \exp[i \int d\alpha~d\beta ~g(\alpha,\beta)
\theta(\alpha,\beta)]
\eqn\eighteen
$$
Because $W(\alpha,\beta)$ and $g(\alpha,\beta)$ satisfy an identical
algebra it follows that the single-particle operator $g^+(\theta(t))
i\partial_t g(\theta(t))$ has an expansion in $g(\alpha,\beta)$ with
coefficients identical to $C_{\alpha\beta}$ in \seventeen :
$$
g^+(\theta(t)) i\partial_t g(\theta(t)) = \int d\alpha~d\beta~
C_{\alpha\beta} (\theta(t),\partial_t \theta(t)) g(\alpha,\beta)
\eqn\nineteen
$$
Now using $\langle x|g(\alpha,\beta)|y\rangle = \delta(x-y+\beta)
e^{i\alpha \left({x+y \over 2}\right)}$ and \seven\
it can be easily shown that
$$
\langle F_0|W(\alpha,\beta)|F_0\rangle = tr(g(\alpha,\beta)\phi_0)
\eqn\twenty
$$
where we have defined
$$
\langle F_0|\Phi|F_0\rangle \equiv \phi_0
\eqn\twentyone
$$
{}From \seventeen, \nineteen\ and \twenty\ we then get
$$
\langle F_0|u^+(\theta(t)) i\partial_\epsilon u(\theta(t))|F_0\rangle =
tr(\phi_0 g^+(\theta(t)) i\partial_t g(\theta(t)))
\eqn\twentytwo
$$

Let us now evaluate the second term in \fifteen.  Since the
single-particle hamiltonian $h$ is an element of the $W_\infty$ algebra we
can expand it in the basis $g(\alpha,\beta)$ as $h \equiv \int
d\alpha~d\beta ~h_{\alpha\beta} g(\alpha,\beta)$.  This implies that the
fermion field theory hamiltonian $H$ has the expansion $H = \int
d\alpha~d\beta \allowbreak ~h_{\alpha\beta} W(\alpha,\beta)$, where we
have used \seven\
to set $tr(g(\alpha,\beta)\Phi)$ equal to $W(\alpha,\beta)$.  We may now
write the second term in \fifteen\ as
$$
\eqalign{
\langle F_{\theta(t)}|H|F_{\theta(t)}\rangle &= \int d\alpha~d\beta~
h_{\alpha\beta} \langle F_0|u^+(\theta(t)) W(\alpha,\beta)
u(\theta(t))|F_0\rangle \cr &
= \int d\alpha~d\beta ~h_{\alpha\beta} \int d\alpha' ~d\beta'
{}~C_{\alpha\beta,\alpha'\beta'} (\theta(t)) \langle
F_0|W(\alpha',\beta')|F_0\rangle \cr &
= \int d\alpha~d\beta~h_{\alpha\beta} \int d\alpha' ~d\beta'~
C_{\alpha\beta,\alpha'\beta'} (\theta(t)) tr(g(\alpha',\beta')\phi_0)}
$$
where in the second step above we have used that $u^+Wu$ is an element of
$W_\infty$ algebra to reexpand it in $W(\alpha,\beta)$ and in the last
step we have used \twenty .  Denoting the single-particle representative
of $u(\theta(t))$ by $g(\theta(t))$ as before, and using in the above an
argument similar to the one given for the identity of coefficients in
\seventeen\ and \nineteen , we get
$$
\eqalign{
\langle F_{\theta(t)} |H|F_{\theta(t)}\rangle &= \int
d\alpha~d\beta~h_{\alpha\beta} tr(g^+(\theta(t)) g(\alpha,\beta)
g(\theta(t)) \phi_0) \cr &
\equiv tr(g^+(\theta(t)) h g(\theta(t)) \phi_0)}
\eqn\twentythree
$$
Putting together \fourteen $-$ \sixteen, \twentytwo\ and \twentythree\
we get
for the short time kernel
$$
K_{t+\epsilon,\epsilon} = \exp [i \epsilon tr\{\phi_0(g^+ i\partial_t g
+ g^+hg)\}]
$$
Hence the finite time kernal is
$$
K = \int \prod_\epsilon d\mu(g(\theta(t))) \exp[i \int dt~tr\{\phi_0
(g^+i\partial_t g + g^+ h g)\}]
\eqn\twentyfour
$$
where $d\mu(g(\theta(t)))$ is an appropriate measure over the coset ${\cal
G}W_\infty/H$.  The path-integral in \twentyfour\ had earlier been
heuristically argued for in [\DAS].

Let us now make contact with the boson action and path-integral measure
given in ref. [\DMW], which we write below (see eqns. (44)
$-$ (46) of this ref.):
$$
\tilde K = \int \prod_t d\mu(\phi_t) \exp i S[\phi]
\eqn\twentyfive
$$
$$
S[\phi] = i \int ds~dt~tr(\phi[\partial_t\phi,\partial_s\phi] + \int
dt~tr(\phi h)
\eqn\twentysix
$$
$$
d\mu (\phi) = \delta (tr~\phi - N) \prod_{x,y} \delta(\phi^2_{xy} -
\phi_{xy}) \prod_{x,y} d\phi_{xy}.
\eqn\twentyseven
$$
In the above, $\phi$ is a hermitian matrix with elements $\phi_{xy} =
\langle x|\phi|y\rangle$.  Also, $\phi(t,s)$ is an extension of $\phi(t)$
such that for $-\infty < t < \infty$ we have $-\infty < s \leq 0$ and the
boundary conditions $\phi(t,s)\big|_{s=0} = \phi(t)$ and
$\phi(t,s)\big|_{s=-\infty} =$ time-independent constant matrix.  We note
that if we set $\phi = g^+ \phi_0 g$, $g \in {\cal G} W_\infty$ and
$\phi_0$ fixed by \twentyone , then the action \twentysix\ and the measure
\twentyseven\ reduce to that appearing in \twentyfour .  The reason for the
measures being identical is that if we fix $\phi_0$ then the integration
in \twentyfive\ is only over the coset obtained by modding out ${\cal
G}W_\infty$ by that subgroup which commutes with $\phi_0$ (i.e. satisfies
for all elements $v$, $v^+ \phi_0 v = \phi_0$).  But this is precisely the
coset ${\cal G}W_\infty/H$ over which the integration in \twentyfour\ is
done.  Let us prove this.  Consider the definition \twentyone\ of $\phi_0$.
Let $v$ be an element of $H$ and let $V$ be its representative in the
fermion fock space.  Then, using $V|F_0\rangle = |F_0\rangle$, for $V \in
H$, we get
$$
\eqalign{
\phi_0 &\equiv \langle F_0|\Phi|F_0\rangle \cr &
= \langle F_0|V^+\Phi V|F_0\rangle \cr &
= v \phi_0 v^+}
$$
The last step follows from arguments similar to those used in deriving
\twentytwo\ and \twentythree .  Thus the two cosets are the same.

For complete identity of the path-integrals in \twentyfour\ and
\twentyfive\
we must, then, explain what restricts the integration over hermitian
matrices in \twentyfive\ to only over the $W_\infty$ ``angles''.  To see
this let us make the change of variables in \twentyfive\ to the ``angles''
and eigenvalues of the hermitian matrix $\phi$:
$$
\phi = g V^0 g^+, ~~~g \in {\cal G}W_\infty, ~~~V^0 ~{\rm diagonal}
\eqn\twentyeight
$$
Putting in \twentysix\ it is easy to show that
$$
S[\phi] = S[g,V^0] = \int dt~tr \{V^0 (g^+i \partial_t g + g^+ h g)\}
\eqn\twentynine
$$
The measure changes to
$$
d\mu (\phi) = \delta (\int d\nu V^0_\nu - N) \prod_\nu [d V^0_\nu
\delta(V^{0^2}_\nu - V^0_\nu)] J(V^0) d\mu(g)
\eqn\thirty
$$
Here $d\mu(g)$ is the measure over the coset obtained by modding out the
subgroup from ${\cal G}W_\infty$ that commutes with $V^0$ and the label
$\nu$ displays the basis in which $\phi$ is diagonal (typically this is
the energy basis).  $J(V^0)$ is the Jacobian of change of variable and
depends only on the eigenvalue matrix $V^0$.  The important point to note
is that the $V^0$ integration can be restricted to a single instant of
time.  This is because the $\delta$-function imposing $\phi^2 = \phi$
implies that the eigenvalues of $\phi$ are only 0 and 1.  Because of the
other $\delta$-function, there are always precisely $N$ number of ones
(fermion number conservation).  This means that the diagonal matrix $V^0$
at all times has $1$ in $N$ number of places and the rest zeros.
Time-dependence can come only in shuffling of the positions of these zeros
and ones.  Since that is achieved by a ${\cal G}W_\infty$ Weyl
transformation, it is already included in the ``angle'' integration.  Thus
the integration over $V^0$ may be restricted to a single instant of time.
Finally, let us discuss the last ingredient needed to completely fix the
functional integral in \twentyfive , namely, a boundary condition on
$\phi(t)$.  This may be given in the form of its value at, say, infinite
past.  For example, a complete specification would be to set
$$
\phi(t)\big|_{t \rightarrow -\infty} = \phi_0
\eqn\thirtyone
$$
where $\phi_0$, defined by \twentyone , corresponds to fermi vacuum.
Clearly, other specifications correspond to different fillings of energy
levels i.e. to excited states.  Hence the complete equivalence of
\twentyfour\ and \twentyfive$-$\twentyseven\ requires specifying a boundary
condition on $\phi(t)$ in \twentyfive\ (in particular, in this case this is
\thirtyone ).

This completes the {\it proof} of the bosonization.  This was earlier done
by us
using the method of co-adjoint orbits of $W_\infty$ [\DMW].  In fact the
$\delta$-functions in the measure in \twentyseven\ and \thirty\ specify
the co-adjoint
orbit of $W_\infty$ corresponding to the representation in terms of $N$,
non-relativistic fermions.

\bigskip

\noindent 3.{\bf Path integral in terms of phase space fluid
density}:

\nobreak
In this section we will reexpress the action \twentysix\ in terms of phase
space density of fermions since it is in terms of this variable that the
semiclassical picture of bosonization in terms of a fermi fluid emerges.
In terms of the fermionic  variables of the action \one, the phase space
density opearator is defined as
$$ \widehat {\cal U} (p,q,t) = \int dx\, \psi^\dagger (q-x/2,t) e^{-ipx}
\psi(q+x/2, t)
\eqn\thirtytwoa $$
We shall  denote its expectation value in a ${\cal G}W_\infty$ coherent
state as $u(p,q,t)$. We also introduce a fourier transform of $u(p,q,t)$:
$$
\tilde u(\alpha,\beta,t) = \int {dp \over 2\pi} {dq \over 2\pi} e^{i(p\beta
- q\alpha)} u(p,q,t)
\eqn\thirtytwob
$$
$\tilde u(\alpha, \beta, t)$ is essentially the expectation value in a
${\cal G}W_\infty$ coherent state of the generator $W(\alpha, \beta, t)$
of $W_\infty$ algebra.

Consider now the expansion of $\phi(\hat x,\hat p)$,
which enters the path integral \twentysix, in terms of a basis for
$W_\infty$ algebra in the single-particle Hilbert  space:
$$
\phi(\hat x,\hat p,t) = \int d\alpha d\beta g(\alpha,\beta) \tilde
u(\alpha,\beta,t)
\eqn\thirtytwo
$$
This expansion is clearly valid since $\phi$ may be thought of as the
expectation value in a ${\cal G}W_\infty$ coherent state of the operator
$\Phi$. In fact \thirtytwob\ and \thirtytwo\ define
the Weyl ordering of $\phi(\hat
x,\hat p)$ corresponding to the classical function $u(p,q,t)$.

In order to express the action \twentysix\
in terms of $u(p,q,t)$, we state a lemma
due to Moyal:

\noindent Lemma (Moyal):

\nobreak
Given two classical functions $f_1(p,q)$ and $f_2(p,q)$ and their
corresponding Weyl ordered operators $\hat f_1(\hat x,\hat p)$ and $\hat
f_2(\hat x,\hat p)$, the classical function corresponding to the
commutator $[\hat f_1,\hat f_2]$ is the fourier transform of the Moyal
bracket,
$$
\eqalign{&
[\hat f_1,\hat f_2] = \int d\alpha d\beta \hat g(\alpha,\beta)
\widetilde{\{f_1,f_2\}}_{MB} (\alpha,\beta) \cr &
\{f_1,f_2\}_{MB} =\big[2 \sin {1 \over 2} (\partial_p \partial_{q'} -
\partial_{p'} \partial_q) (f_1 (p,q) f_2(p',q'))\big]_{p'=p,q'=q}}
\eqn\atwentytwo
$$
Note that  in the second equation above the first term in the expansion of
$\sin(\partial_p\partial_{q'}-\partial_{p'}\partial_q)$ is just the
Poisson bracket.  In \atwentytwo\ the trace identity $tr[A,B] = 0$
is implicit.  Restricting to such operators is equivalent to requiring
$\int\int dp~dq\{a(p,q),b(p,q)\}_{MB} = 0$, for the corresponding
classical functions.  This can be achieved by requiring the boundary
condition that $a(p,q)$ and $b(p,q)$ are constant as $p,q \rightarrow \infty$.

Using \thirtytwob\ , \thirtytwo\ and \atwentytwo\ we
can easily see that the action \twentysix\ becomes,
$$
\eqalign{
S[u] = &\int ds dt \int {dp dq \over 2\pi } u(p,q,t,s)
\left\{\partial_s u(p,q,t,s),\partial_t
u(p,q,t,s)\right\}_{MB} \cr &
+ \int dt \int {dp dq \over 2\pi } h(p,q) u(p,q,t)}
\eqn\atwentythree
$$
and the measure \twentyseven\ becomes
$$
d\mu (u) = \delta\left(\int {dp dq \over 2\pi }
u(p,q,t) - N\right)
\prod_{p,q} \left[\delta(C(p,q,t)) du(p,q,t)\right].
$$
This implies the constraints,
$$
C(p,q,t) \equiv \big[
\cos {1 \over 2} (\partial_p \partial_{q'} - \partial_{p'}
\partial_q) (u(p,q,t) u(p',q',t))\big]_{p'=p,q'=q} -
u(p,q,t) = 0
\eqn\atwentyfour
$$
$$
\int {dp dq \over 2\pi } u(p,q,t) = N
\eqn\atwentyfive
$$
To derive the equation of motion from \atwentythree\
we make a variation
$\delta u(p,q,t) = \{\epsilon,u\}_{MB}$ that preserves the
constraints \atwentyfour, \atwentyfive. This gives
$${\partial \over \partial t} u(p,q,t) + \{h,u\}_{MB}
(p,q,t) = 0.
$$
This is the `quantum' version of Liouville's equation.  It
is worth mentioning that if $h = {1\over2} (p^2-q^2)$, the Moyal bracket
equals the Poisson bracket and we get
$$
\partial_t u + (p\partial_q + q\partial_p)u = 0
\eqn\atwentysix
$$

\bigskip

\noindent 4.{\bf  Weak coupling (semiclassical) limit}:

\nobreak
We are now in a position to discuss the weak coupling limit.
We shall restrict our discussion to the single-particle hamiltonian
$h = {1\over2} (p^2-q^2)$ which is obtained in the double scaling limit of
$c=1$ matrix model. The discussion can be easily generalized to other
cases. As has been
explained in detail in [\DMW], the semiclassical limit of this theory is
obtained by expanding in a power series  both the `sine' in the Moyal
Bracket \atwentytwo\ and the `cosine'
in the constraint  \atwentyfour\ and
retaining only the first term in both cases.
In this limit the constraint
simplifies to $u^2(p,q,t) = u(p,q,t)$, implying that the
configurations that
enter the path integral are characteristic functions corresponding to
regions in phase space.  The only dynamical part of a characteristic
function is the boundary of the region and
the dynamics consists of changes of the boundary preserving the area
enclosed.
One can indeed give a precise
description of this using the classical limit of $W_\infty$ algebra which
is the algebra of
classical canonical transformations in two dimensions.
We refer the reader for details to ref. [\DMW].

Let us now see under what precise approximations
does collective field theory emerge from
the above semiclassical limit.
Consider defining the moments of $u(p,q,t)$,
$$
\eqalign{&
\rho(q,t) \equiv {\tilde \rho(q,t) \over 2\pi} =
\int^{+\infty}_{-\infty} {dp \over 2\pi} u(p,q,t) \cr &
\pi(q,t) \rho(q,t) = \int^{+\infty}_{-\infty} {dp \over 2\pi} p
u(p,q,t) \cr &
\pi_2(q,t) \rho(q,t) = \int^{+\infty}_{-\infty} {dp \over 2\pi} p^2
u(p,q,t) ~~~~~~~~~{\rm ,etc.}}
\eqn\atwentyeight
$$
The equations of motion for the moments $\rho(q,t),\pi(q,t),\pi_2(q,t)$
etc. can be derived from the equation of motion \atwentysix,
$$
\eqalign{&
\partial_t\tilde \rho + \partial_q (\tilde \rho \pi) = 0 \cr &
\partial_t \pi = \partial_q \left({\pi^2 \over 2} + {q^2 \over 2} -
\pi_2\right) + {\partial_q \rho \over \rho} (\pi^2 - \pi_2) \cr &
{\rm etc.}}
\eqn\atwentynine
$$
Furthermore, the constraint \atwentyfour\ implies certain relations among
the moments.  In the semiclassical limit
the ground state is described by
$$
u_0(p,q) = \theta \left(\mu - {p^2 - q^2 \over 2}\right) =
\theta\left[(\sqrt{q^2 + 2\mu} - p) (p + \sqrt{q^2+2\mu})\right]
$$
where the curve ${p^2 - q^2 \over 2} = \mu$ defines the fermi surface.
Collective field theory is defined by parametrizing $u(p,q,t)$ near $u_0
(p,q)$ by [\POL]
$$
u(p,q,t) = \theta\left[(p_+(q,t) - p) (p - p_-(q,t))\right]
\eqn\athirty
$$
where
$p_+(q,t)$ and $p_-(q,t)$ are such that $|p_\pm (q,t) - \sqrt{q^2+2\mu}|$ is
small.
In other words the collective field
theory approximation to the semiclassical limit is described by
those low energy excitations
of the fermi fluid near the fermi surface  which
are described by a curve
quadratic in $p$.  This assumption, together with the semiclassical
constraint $u^2=u$,  leads to  specific
relations between the moments  $\rho,\pi,\pi_2$ etc.  In particular,
we have
$$
\pi_2 = \pi^2 + {1 \over 12} \tilde\rho^2
\eqn\athirtyone
$$
Substituting this in \atwentynine\ we obtain the equations of collective
field theory:
$$
\eqalign{&
\partial_t \tilde\rho + \partial_q (\pi\tilde\rho) = 0 \cr &
\partial_t \pi + \pi \partial_q \pi = -\partial_q\left(-{q^2 \over 2} +
{\tilde\rho^2 \over 8}\right)}
\eqn\athirtytwo
$$
It is easy to see that a generic boundary ({\it i.e.} not necessarily
quadratic in $p$) violates the above equations. In fact, a generic
boundary is not even described in terms of the first two moments. We have
shown in [\DMWTIME] in an explicit example how this can result in physical
quantities having different values from those calculated from collective
field theory {\it even} in the semiclassical limit.

We thus see that if we restrict ourselves to those shapes of the fermi
fluid that have a quadratic profile then the semiclassical approximation
reduces to collective field theory.

To summarize, in this paper we have presented a first-principles proof of
bosonization of noninteracting, nonrelativistic fermions in one space
dimension and obtained the bosonic action. In the semiclassical limit the
bosonized theory reduces to the dynamics of area-preserving fermi fluid
profiles in the phase space. Restricting to quadratic profiles gives the
collective field theory.

\refout
\end